\documentclass{article}

\def\abstract#1{\vskip 7mm 
        \begin{center}{\large Abstract}\par \smallskip
                \begin{minipage}[c]{12cm}
                        \small #1
                \end{minipage}
        \end{center}
}
\def\title#1{\begin{center}{\Large\bf #1}\end{center}}
\def\author#1{\vskip 5mm \begin{center}{#1}\end{center}}
\def\address#1{\begin{center}{\it #1}\end{center}}
\makeatletter
\@ifundefined{lesssim}{}{}
\@ifundefined{gtrsim}{}{}
\def\vereq#1#2{\lower3pt\vbox{\baselineskip1.5pt \lineskip1.5pt
\ialign{$\m@th#1\hfill##\hfil$\crcr#2\crcr\sim\crcr}}}
\makeatother

\begin{document}

\title{%
  Multi-graviton theory in vierbein formalism
}
\author{%
  Teruki Hanada,\footnote{E-mail : k004wa@yamaguchi-u.ac.jp}
  Kazuhiko Shinoda\footnote{E-mail : k009vc@yamaguchi-u.ac.jp}
  and
  Kiyoshi Shiraishi\footnote{E-mail : shiraish@yamaguchi-u.ac.jp}
}
\address{%
  Graduate School of Science and Engineering, Yamaguchi University,
   Yoshida, Yamaguchi-shi, Yamaguchi 753--8512, Japan
}

\abstract{
Recently, multi-graviton theory on a simple closed circuit graph
corresponding to the $S^1$ compactification of the Kaluza-Klein (KK) theory has
been considered. 
In the present paper, we extend this theory to that on a general graph and study what
modes of particles are included. 
Furthermore, 
we generalize it in a possible non-linear theory based on the vierbein formalism
and study cosmological solutions.%
}

\section{Introduction}
Both astronomical and cosmological data seem to require the presence of yet directly
 undetected dark matter and 
dark energy in the universe. The necessity for these mysterious components occurs at
distances where the gravitational 
interaction is not understood sufficiently. This suspicious coincidence inspires a 
search for modifications of the 
general relativity at large distances.

It is important for understanding cosmology and unification to study massive and 
multi-graviton theory.
In the linear theory, gravitons have a Fierz-Pauli (FP) type mass~\cite{ref1}.
But there is an ambiguity in its nonlinear generalization.
We study thus far the linear multi-graviton theory on a circle corresponding to
$S^1$ compactification of 
KK theory with dimensional deconstruction~\cite{ref2}. This model is an extended
version  of Hamamoto's model~\cite{ref3}.

In this article,  we construct the FP Lagrangian on a general graph and investigate
what modes of particles are included.
Furthermore, we extend it to a nonlinear theory based on the vierbein formalism.

\section{FP on a graph}
We consider the matrix representation of the graph theory.\footnote{Please see
\cite{refjmp} for a review of application of graph theory to field theory.} A graph $G$
is a pair of $V$ and $E$, where
$V$ is a set of vertices while
$E$ is a set of edges. An edge connects two vertices; two vertices located at the
ends of an edge $e$ are denoted as $o(e)$ and $t(e)$. Then, we introduce two matrices,
an incidence matrix and a graph Laplacian, associated with a specific graph. The
incidence matrix represents the condition of connection or structure of a graph,  and 
the graph Laplacian $\Delta$ can be obtained by  $E E^T$. By use of these matrices, 
a quadratic form of vectors $a^T \Delta a(=a^T E E^T a)$ can be written as a sum of
$(a_i-a_j)^2$. If all $a_i~(i=1, 2, \dots, \#V)$, components of $a$, take the same
value, $E^T a=0$ and then
$\Delta a=0$.

So, we consider the Lagrangian for a massive graviton $h^v_{\mu\nu}$ on each vertex
with the Stueckelberg vector field $A^e_{\mu}$
on each edge and a scalar field $\phi^v$ on each vertex:
\begin{eqnarray*}
L_m \hspace{-.5em}
&=& \hspace{-.5em}
L_0-\frac{m^2}{2}\sum_{v\in V}\left[h^{v\mu\nu}(EE^T h_{\mu\nu})^v
-h^v(EE^Th)^v\right]\\
& &\hspace{3cm}-2\sum_{v\in V}\left[m(E A_\mu)^v+\partial_\mu \phi^v\right]
(\partial_\nu h^{v\mu\nu}-\partial^\mu h^v)-\frac{1}{2}\sum_{e\in E}
\left(\partial_\mu A^e_\nu -\partial_\nu A^e_\mu\right)^2,
\end{eqnarray*}
where $L_0$ is the linearized Einstein-Hilbert Lagrangian:
\[
L_0=\sum_{v\in V}\left[-\frac{1}{2}\partial_\lambda h^v_{\mu\nu}\partial^\lambda h^{v\mu\nu}
+\partial_\lambda h^{v\lambda}_{\ \ \mu}\partial_\nu h^{v\nu\mu}
-\partial_\mu h^{v\mu\nu}\partial_\nu h^v-\frac{1}{2}\partial_\lambda h^v 
\partial^\lambda h^v\right],
\]
and $h^v\equiv\eta^{\mu\nu}h^v_{\mu\nu}$.

This action is invariant under the following transformations:
\[
h^v_{\mu\nu}\rightarrow h^v_{\mu\nu}+\partial_\mu \xi^v_\nu +\partial_\nu \xi^v_{\mu},
\quad A^e_\mu \rightarrow A^e_\mu +m(E^T \xi_\mu)^e -\partial_\mu \zeta^e,
\quad \phi^v\rightarrow \phi^v + m(E\zeta)^v,
\]
where $\xi^v$ and $\zeta^e$ are parameters on each vertex and each edge respectively.

Suppose the following gauge fixing terms:
\[
L_{gf}=-\sum_{v\in V}\Big[\partial_\nu h^{v\mu\nu}-\frac{1}{2}\partial^\mu h^v-m(EA^\mu)^v
-\partial^\mu \phi^v\Big]^2 -\sum_{e\in E}\Big[\partial_\mu A^{e\mu}-\frac{m}{2}(E^Th)^e
-m(E^T\phi)^e\Big]^2,
\]
then, the gauge-fixed Lagrangian becomes 
\[L_m+L_{gf}=\frac{1}{2}H^{\mu\nu}(\partial^2-m^2EE^T)\Big(H_{\mu\nu}-\frac{1}{2}H\eta_{\mu\nu}\Big)
+A^\mu(\partial^2-m^2E^TE)A_\mu+3\phi(\partial^2-m^2EE^T)\phi,\]
where $H_{\mu\nu}=h_{\mu\nu}+\phi\eta_{\mu\nu}$. Here the indices $v$ and $e$, and
the notion of sum over them are omitted.

\section{Dimensional deconstruction}
It is assumed that we put fields on vertices or a edges. An idea that there are four
dimensional
 fields on the sites (vertices) and links (edges), dimensional deconstruction, is
introduced by Arkani-Hamed {\it et al}.~\cite{ref4,ref5}. In this scheme, the square
 of
mass matrix is proportional to the Laplacian of the associated graph.
 
In the case of a cycle graph (a `closed circuit') with $N$ sites ($C_N$), when $N$
becomes large, the model on the graph corresponds with the five-dimensional theory
with $S^1$ compactification. In other words, the mass scale of the model
$f$ over
$N$ correspond to the inverse of the compactification radius:
\[M_\ell^2=4f^2(\sin\pi\ell/N)^2\quad \rightarrow\quad  M_\ell^2=(2\pi\ell/L)^2,
\qquad (f/N \rightarrow 1/L)\,.\]

For a cycle graph, the linear graviton model presented in the previous section
coincides with the model proposed in \cite{ref2}. The model is
a most general linear graviton theory on a generic graph.

\section{Multi-graviton theory on a general graph}
For this model, we investigate what modes of particles are contained. Although any
graph is valid for the model, here we consider
 two examples, a cycle graph $C_N$ and a path graph $P_N$. In the case of the cycle
graph $C_N$  ($\#V=N,$ $\#E=N$),  $N-1$ massive spin two's, a massless spin two,
$N-1$ massive vectors, a massless vector, $N-1$ massive scalars, and
a massless scalar seem to be included, as seen from the gauge-fixed Lagrangian.
The mass spectra of different spin fields are the same, up to zero modes.
This is due to the fact that eigenvalues of $EE^T$ and ones of $E^TE$ are the same
except for zero eigenvalues. 

However,   $N-1$ massive spin two, a massless spin two, a
massless vector,  and a massless scalar are left physically, because massive vectors
and massive scalars are absorbed by massive spin two fields to form massive gravitons
with five degrees of freedom each.
 
 Similarly, in the case of the path graph $P_N$ ($\#V=N,$ $\#E=N-1$), $N-1$ massive
spin two's, a massless spin two, and a massless scalar is left physically, the massless
vector mode is absent.

The limits of $N$ to infinity in the cases of
$C_N$ and $P_N$ realize the KK theory with $S^1$ and $S^1/Z_2$ compactification,
respectively.

\section{Nonlinear generalization}
Now we will consider a nonlinear extension of the linear theory. Following 
Nibbelink {\it et al}.~\cite{ref6,ref7}, we introduce a useful `tool':
\[\langle ABCD \rangle\equiv -\varepsilon_{abcd}\varepsilon^{\mu\nu\rho\sigma}A^a_\mu
B^b_\nu C^c_\rho D^d_\sigma,\] where $\varepsilon$ is the totally antisymmetric
tensor. Using this expression, we have the Einstein-Hilbert term replacing $A$ and $B$
by vierbeins and $C$ and $D$ by the curvature 2-form. In addition, because fourth power
of vierbein in the angle  bracket
 is equal to the determinant of vierbeins ($\langle eeee\rangle=\langle
e^4\rangle=|e|$), this expression means that the Einstein-Hilbert term and the
cosmological term  have the same structure.

We now assume that the following term is assigned for each edge of a graph:
\[\langle (e_1e_1 -e_2e_2)^2 \rangle,\]
where $e_1$ and $e_2$ are vierbeins at two ends of one edge. Note that this term has a
reflection symmetry $e\leftrightarrow -e$  at each vertex and an exchange symmetry $e_1
\leftrightarrow e_2$ at each edge.

In the weak field limit, {\it i.e.} $e_1=\eta +f_1$, $e_2=\eta+ f_2$,
\[\langle(e_1e_1-e_2e_2)^2\rangle=
8\left(\left(\left[f_1\right]-\left[f_2\right]\right)^2
-\left[\left(f_1-f_2\right)^2\right]\right)+O(f^3)\,,\]
where $\eta$ is the Minkowski metric, and $[f]={\rm tr}f$ for notational simplicity.
This quadratic term corresponds to  FP mass term.\footnote{It is known that the
asymmetric part of $f$ can be omitted~\cite{refbiz}.}

On the other hand, the Einstein-Hilbert term $\frac{1}{2}|e|R$ contains the kinetic
terms of a graviton in the lowest order up to the total derivative: 
\[\frac{1}{2}|e|R
=
-\frac{1}{2}\partial_\lambda f_{\mu\nu}\partial^\lambda f^{\mu\nu}
+\partial_\lambda f^\lambda_{\ \mu}\partial_\nu f^{\nu\mu}\\-\partial_\mu f^{\mu\nu}\partial_\nu f
-\frac{1}{2}\partial_\lambda f\partial^\lambda f +O(f^3)\,,\]
and $\frac{1}{2}R$ contains the following terms in the first order:
\[\frac{1}{2}R
=
-\partial^\lambda \partial_\lambda f+\partial_\mu 
\partial_\nu f^{\mu\nu}+O(f^2)\,.\]

In the case of a tree graph (a graph with no closed circuit---the path graph $P_N$ is
a tree graph, for example), we have the nonlinear Lagrangian of multi-graviton theory 
without higher derivertive and non-local terms,
\[L_M=\frac{1}{2}\exp\Phi\sum_{v\in V}\left|e^v\right|R^v+M^2\sum_{e\in E}
\left\langle\left(e_{o(e)}e_{o(e)}-e_{t(e)}e_{t(e)}\right)^2\right\rangle,\]
where $M^2\equiv m^2/16$.
The scalar zero-mode field $\Phi$ can be identified as $\phi_1=\phi_2=\cdots
=\Phi$.

\section{Cosmological solution}
We will derive a cosmological solution of our model on a tree graph with $N$ vertices.
We assume
$L=L_M+L_\Lambda$, where
\[L_\Lambda=\exp(a\Phi)\sum_{v\in V}\left| e^v\right|\Lambda^v\,.
\] 
Here, $L_\Lambda$ represents the simplest effects of matters and $a$ is a coupling
constant.  Suppose that each metric is homogeneous, isotropic, and flat, {\it i.e.}
\[
ds_k^2=-B_k^2(t)dt^2+A_k^2(t)d\vec{x}^2\,.
\]
Moreover, if we assume
\[\quad A_k(t)=\alpha_kA(t), \quad B_k(t)=\alpha_k, \quad \frac{\dot{A}}{A}=H\,,\]
and $\Lambda_1=\Lambda_2=\cdots 
=\Lambda_N=\Lambda=\lambda_{(\ell)} M^2$, we obtain the following solution;
\[\alpha_k^2=\frac{3H^2}{\Lambda}\left(1+\sqrt{2}\frac{\sqrt{2-a}}{\sqrt{a}}
v_k^{(\ell)}\right),\qquad \left(\frac{4}{3}<a<2\right),\qquad \Phi\equiv 0,\]
where $\{\lambda_{(\ell)},v_k^{(\ell)}\}$ are the nonzero eigenvalues and the
components of eigenvectors.
Note that this `synchronized' solution is not a general solution but a special one.

Interestingly enough, in the case of the path graph, we have other
solutions of Randall-Sundrum type by tuning the value of $\Lambda_1$ and
$\Lambda_N$.

\section{Summary and prospects}
We have studied the simple theory of multi-graviton, and have shown a cosmological
solution. We should investigate more plausible solution for classical as well
as quantum cosmology, including usual matter.

To this end, we should study the graviton coupling to various matter fields.
At the same time, we expect that the Higgs-like mechanism on gravity might be developed
by pursuing complicated, non-minimal interactions with matters.
Incorporating SUSY (SUGRA) is also of much interest.

Permitting higher-derivative terms and non-local terms in the action will
bring more possibilities to the completion of nonlinearity and be worth studying
still.

From the mathematical point of view, it is interesting to construct models
with the use of generic graphs, such as weighted graphs, fractals, and so on.

\section*{Acknowledgements}
The authors would like to thank N.~Kan for useful comments, and also the organizers of JGRG17.

\end{document}